\documentstyle[aps,psfig]{revtex}

\def\relaz#1#2{\mathrel{\mathop{\kern 0pt#1}\limits_{#2}}}

\def\bS{{\mbox{\boldmath $S$}}}

\def\bi{{\bf i}}
\def\bj{{\bf j}}

\begin{document}
\wideabs{
\title{The quantum Heisenberg antiferromagnet on the square lattice}

\author{Alessandro Cuccoli, Valerio Tognetti and Paola Verrucchi}
\address{Dipartimento di Fisica dell'Universit\`a di Firenze
        and Istituto Nazionale di Fisica della Materia (INFM),
        \\ Largo E. Fermi~2, I-50125 Firenze, Italy}

\author{Ruggero Vaia}
\address{Istituto di Elettronica Quantistica
        del Consiglio Nazionale delle Ricerche,
        via Panciatichi~56/30, I-50127 Firenze, Italy,
        \\ and Istituto Nazionale di Fisica della Materia (INFM)}


\maketitle

\begin{abstract}
The pure-quantum self-consistent harmonic approximation, a
semiclassical method based on the path-integral formulation of
quantum statistical mechanics, is applied to the study of the
thermodynamic behaviour of the quantum Heisenberg antiferromagnet on
the square lattice (QHAF).  Results for various properties are
obtained for different values of the spin and successfully compared
with experimental data.
\end{abstract}

}

We consider the quantum Heisenberg antiferromagnet on the square lattice
(QHAF), whose Hamiltonian reads
\begin{equation}
 \hat{\cal{H}}= J\sum_{<{\bf{i}}{\bf{j}}>}
 \hat {\mbox{\boldmath$S$}}_{\bf{i}} \cdot
 \hat {\mbox{\boldmath$S$}}_{\bf{j}}~;
\label{e.H}
\end{equation}
$J$ is positive, the sum runs over all the couples $<{\bf{i}}{\bf{j}}>$
of neighbouring sites on the square lattice, and the 
quantum operators $\hat{\bS}_\bi$ obey the 
angular momentum commutation relations 
$[S_\bi^\alpha,S_\bj^\beta]=iS_\bi^\gamma\delta_{\bi\bj}
\varepsilon^{\alpha\beta\gamma}$ with $|\hat{\bS}_\bi|^2=S(S+1)$.

Several real compounds are well described, as far as their magnetic
behaviour is concerned, by this model with $S=1/2$ (La$_2$CuO$_4$,
Sr$_2$CuO$_2$Cl$_2$), $S=1$ (La$_2$NiO$_4$, K$_2$NiF$_4$) and $S=5/2$
(KFeF$_4$, Rb$_2$MnF$_4$), and a consequently rich experimental
analysis of the subject has been developed in the last ten years. From
the theoretical point of view, an equally rich reservoir of results,
from both analytical and numerical approaches, is now available;
nevertheless, there are still many open questions, and different
conclusions have been recently drawn by several
authors~\cite{BeardEtal97,KimT98,LeeEtal98}.

To study the QHAF, we have used the pure-quantum self-consistent 
harmonic approximation (PQSCHA)~\cite{CGTVV95}, which 
is a semiclassical method based on the path-integral formulation 
of quantum statistical mechanics.
Its main feature is that of exactly describing 
the classical behaviour and fully take into account 
the linear part of the quantum contribution to the thermodynamics of the 
system, so that the self-consistent harmonic approximation
is only used to handle the pure-quantum nonlinear contribution.
The fundamental goal of the PQSCHA is that of reducing the evaluation of 
quantum statistical averages to classical-like expressions involving
properly renormalized functions, the fundamental one being the effective 
Hamiltonian ${\cal{H}}_{\rm eff}$. 
If $\beta=T^{-1}$, $N$ is the number of lattice sites, 
${\mbox{\boldmath$s$}}_{\bf i}$ is a classical
vector on the unitary sphere  ($|{\mbox{\boldmath$s$}}_{\bf i}|=1$),
and $\int d^{\scriptscriptstyle N}\!{\mbox{\boldmath$s$}}$ 
indicates the phase-space integral for a classical magnetic system,
the quantum statistical average of a physical observable described
by the quantum operator ${\hat{O}}$ turns out to be
$
\langle\hat {\cal O}\rangle=1/{\cal Z}
\int d^{\scriptscriptstyle N}\!{\mbox{\boldmath$s$}}
 ~\widetilde{\cal O}~
 \exp (-\beta {\cal H}_{\rm{eff}})
$, where ${\cal{Z}}{=}\int d^{\scriptscriptstyle N}\!{\mbox{\boldmath$s$}}
 ~\exp(-\beta{\cal{H}}_{\rm{eff}})$ is the partition function. 
Both ${\cal{H}}_{\rm eff}$ and $\widetilde{\cal O}$ depend on $T$ and $S$,
and the determination of their explicit form, starting from
the expression of the original quantum operators, is indeed the core of the 
method~\cite{CTVV9798prb}. 
The effective Hamiltonian for the QHAF, i.e. relative to Eq.(\ref{e.H}),
is
\begin{equation}
 {{\cal H}_{\rm{eff}}\over J\widetilde{S}^2}
 = \theta^4\sum_{<{\bf{i}}{\bf{j}}>} {\mbox{\boldmath$s$}}_{\bf{i}}
 {\cdot} {\mbox{\boldmath$s$}}_{{\bf{j}}} + {\cal G}(t)~,
\label{e.Heff}
\end{equation}
where
$\widetilde{S}=S+1/2$, $t=T/J\widetilde{S}^2$ and 
${\cal{G}}(t)$ is a uniform term that does not affect the evaluation of 
statistical averages. The renormalization coefficient 
$\theta^2=\theta^2(t,S)<1$ is easily evaluated, for any given $t$ and $S$,
by self-consistently solving two coupled equations~\cite{CTVV9798prb}.
>From Eq.(\ref{e.Heff}) we see that the quantum effects leave the simmetry
of the Hamiltonian unchanged and introduce an energy scaling factor 
$\theta^4$, naturally defining the effective classical temperature
\begin{equation}
t_{\rm eff}={t\over \theta^4(t,S)}
\label{e.teff}
\end{equation}
that will enter all the PQSCHA results.
The partition function, for instance, is 
${\cal{Z}}=\exp[{-\beta{\cal{G}}(t)}]{\cal{Z}}_{\rm cl}(t_{\rm eff})$,
where ${\cal{Z}}_{\rm cl}(t_{\rm eff})$ is the partition function
of the classical model at a temperature $t_{\rm eff}$; 
 $O_{\rm cl}(t_{\rm eff})$ will hereafter mean the 
value taken by the quantity $O$ in the classical Heisenberg
antiferromagnet at a temperature $t_{\rm eff}$.

The internal energy per site is easily found to be
$
u(t)=\theta^4(t,S)u_{\rm cl}(t_{\rm eff})~$, while the correlation functions 
$G({\bf{r}})\equiv\langle\hat{\mbox{\boldmath$S$}}_{\bf{i}}
{\cdot}\hat{\mbox{\boldmath$S$}}_{{\bf{i}}+{\bf{r}}}\rangle$, with
${\bf{i}}\equiv(i_1,i_2)$ and 
${\bf{r}}\equiv(r_1,r_2)$ any vector on the square lattice, turn out to be 
\begin{equation}
G({\bf{r}},t)= \widetilde{S}^2 \theta^4_{\bf{r}}
G_{\rm cl}({\bf{r}},t_{\rm eff})~;
\label{e.Gr}
\end{equation}
the renormalization coefficients 
$\theta^2_{\bf{r}}=\theta^2_{\bf{r}}(t,S)$
are such that $\theta^2_{\bf{r}}$ does not depend on $\bf{r}$ for large
$|\bf{r}|$, and $\theta^2_{\bf{r}}=\theta^2$ for $|{\bf{r}}|=1$.
 From Eq.(\ref{e.Gr}), we find the PQSCHA expression for the staggered 
susceptibility $\chi\equiv\sum_{{\bf{r}}}(-)^{r_1+r_2}~G({\bf{r}},t)/3$ 
to be
\begin{equation}
 \chi={1\over 3}\bigg[
 S(S+1) + \widetilde{S}^2 \sum_{{\bf{r}}\neq 0} (-)^{r_1+r_2}
 ~\theta^4_{{\bf{r}}} 
G_{\rm cl}({\bf{r}},t_{\rm eff})\bigg]~.
\label{e.chi}
\end{equation}

The PQSCHA result for the correlation length, defined by
the asymptotic expression $G({\bf{r}})\propto\exp(-|{\bf{r}}|/\xi)$ for large
$|{\bf{r}}|$, is 
$
\xi(t)=\xi_{\rm cl}(t_{\rm eff})$,
meaning that the correlation length of the QHAF at a temperature $t$
equals that of its classical counterpart at a temperature $t_{\rm eff}$.

Once the problem has been reduced, by the PQSCHA, to a 
renormalized classical one,
the ingredients needed to obtain the temperature and spin dependent
thermodynamic properties of the QHAF are the renormalization
coefficients $\theta^2_{\bf{r}}(t,S)$, whose evaluation is a simple matter
of a fraction of second on a standard PC, and the temperature dependence of the
corresponding properties of the classical model, typically obtained 
by classical Monte Carlo simulations~\cite{MC}.

In the following we will focus our attention on the staggered susceptibility
and the correlation length, as experimental data for these quantities are
available for various compounds. Such compounds 
are usually characterized by a crystal structure in which 
the magnetic ions form parallel planes and mainly interact if belonging 
to the same plane; a weak interplanar interaction is responsible for 
a low-temperature 3D transition, and it introduces also an anisotropy term. 
Keimer et al.~\cite{KeimerEtal92} have shown that  
in the classical limit, and to one-loop level, 
the relation between $\xi$ in presence of the anisotropy term,
and $\xi_0$ of the fully isotropic model, is given by
$
\xi=\xi_0/(1-\alpha\xi_0^2)^{1/2}$,
where $\alpha$ is a parameter describing the relative strength of anisotropy;
following Lee et al.~\cite{LeeEtal98} we shall employ
the above formula  (with some refinements~\cite{CTVVmmm98} which lead to 
substitute $\alpha$ with its renormalized counterpart $\alpha_{\rm eff}$ ) 
to compare our PQSCHA results with experimental data.

\begin{figure}[h]
\centerline{\psfig{bbllx=16mm,bblly=75mm,bburx=197mm,bbury=207mm,%
figure=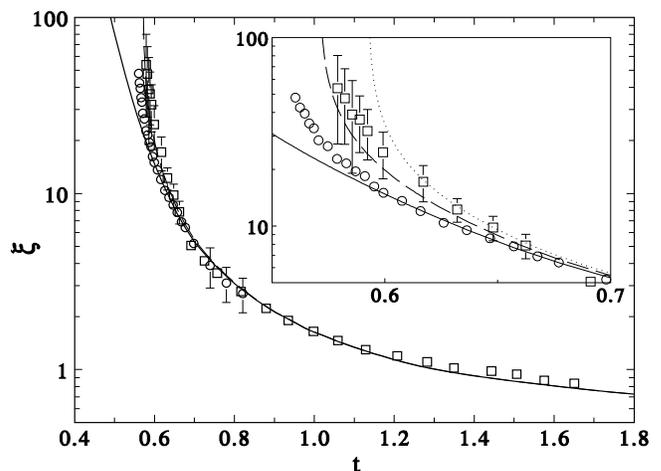,width=85mm,angle=0}}
\caption{Correlation length $\xi$ versus $t$ for $S=5/2$:
Lines are the PQSCHA results for the QHAF with the bare $\alpha$ (dotted),
the renormalized $\alpha_{\rm eff}$ (dashed) and without anisotropy (full).
Experimental data for Rb$_2$MnF$_4$ \protect\cite{LeeEtal98}
(squares) and KFeF$_4$\protect\cite{FultonEtal94}(circles).
\label{f.xi5/2}
}
\end{figure}

\begin{figure}[h]
\centerline{\psfig{bbllx=12mm,bblly=73mm,bburx=194mm,bbury=203mm,%
figure=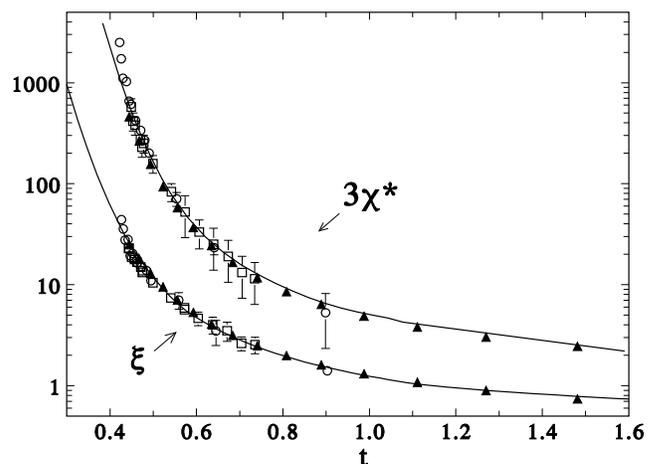,width=85mm,angle=0}}
\caption{Correlation length $\xi$ and 
staggered susceptibility $3\chi^*=3\chi/\widetilde{S}$ 
versus $t$ for $S=1$:
Experimental data for La$_2$NiO$_4$\protect\cite{NakajimaEtal95}(squares),
and K$_2$NiF$_4$\protect\cite{GrevenEtal94}(circles); QMC data (triangles) 
from Ref.\protect\cite{HaradaEtal98}.
\label{f.xichi1}
}
\end{figure}

In Figs.~\ref{f.xi5/2} and~\ref{f.xichi1} we present our results for the correlation length of the QHAF
for $S=5/2$, and for the same quantity and the staggered
susceptibility for $S=1$. For $S=5/2$ the experimental data refer to Rb$_2$MnF$_
4$~
\cite{LeeEtal98} and KFeF$_4$~\cite{FultonEtal94}: 
the anisotropy term is seen to be very well described by the approach 
described above. For $S=1$ QMC~\cite{HaradaEtal98} and experimental data for the
two compounds La$_2$NiO$_4$~\cite{NakajimaEtal95} 
and K$_2$NiF$_4$~\cite{GrevenEtal94} are reported. 

We may thus confidently conclude that the thermodynamic behaviour of the QHAF
is properly described by the PQSCHA, i.e. in term of a renormalized 
classical Heisenberg antiferromagnet. The easy-axis anisotropy,
sometimes crucial to analyse the low-temperature experimental data,
has been considered following Keimer et al.~\cite{KeimerEtal92}, in the 
PQSCHA framework.
Finally, we would like to recall that our results help to shed some
light on the reasons of the failure of the theory based on the
non-linear $\sigma$ model approach~\cite{ChakravartyHN89HasenfratzN91}, 
in describing the QHAF for $S\ge 1$ in the temperature region where 
experimental data are available. 

\vspace{5mm}

\vspace{5mm}

\end{document}